# SEEMS: A Single Event Effects and Muon Spectroscopy facility at the Spallation Neutron Source

Travis J. Williams[1,*], Gregory J. MacDougall[2,+], Bernie W. Riemer[3], Franz X. Gallmeier[4], Clarina R. dela Cruz[1] and Despina Louca[5]

[1]Neutron Scattering Division, Oak Ridge National Laboratory, Oak Ridge, TN 37831
[2]Materials Research Laboratory and Department of Physics, University of Illinois at Urbana-Champaign, Urbana, IL 61801
[3]SNS Upgrades Office, Oak Ridge National Laboratory, Oak Ridge, TN 37831
[4]Neutron Technologies Division, Oak Ridge National Laboratory, Oak Ridge, TN 37831
[5]Department of Physics, University of Virginia, Charlottesville, VA 22904

This manuscript outlines a concept that would leverage the existing proton accelerator at the Spallation Neutron Source (SNS) of Oak Ridge National Laboratory to enable transformative science via one world-class facility serving two missions: Single Event Effects (SEE) and Muon Spectroscopy (µSR). The µSR portion would deliver the world's highest flux and highest resolution pulsed muon beams for materials characterization purposes, with precision and capabilities well beyond comparable facilities. The SEE capabilities deliver neutron, proton and muon beams for aerospace industries that are facing an impending challenge to certify equipment for safe and reliable behavior under bombardment from atmospheric radiation originating from cosmic and solar rays. With negligible impact on the primary neutron scattering mission of the SNS, the proposed facility will have enormous benefit for science and industry alike. We have designated this facility 'SEEMS'.

**1. Introduction and background**

Muon Spin Rotation/Relaxation/Resonance (µSR) is a spectroscopic tool using beams of spin-polarized muons at particle accelerators to perform world-class scientific experiments in condensed matter physics, engineering, energy materials and chemistry. µSR has a long, well-established history of high-profile scientific publications in all these areas [1,2,3,4,5,6,7] and dedicated µSR facilities currently exist at 4 particle accelerators world-wide. Despite a history of pioneering developments of the µSR technique and a growing user base, there is currently no facility in the United States capable of performing these experiments.

Single Event Effects (SEE) are a collection of errors and faults in semiconductor and other electronic devices that result from ionizing radiation – even at low levels [8,9]. The ever-increasing application of complex computing systems in aircraft, ground transportation, autonomous vehicles and mission critical computing systems is providing a greater chance for SEE events to lead to system faults with critical safety or mission reliability impact [9]. As such, industry standards boards and regulatory agencies are beginning to require SEE assessments which are built upon irradiation test data. These changes will lead to significant increases in demand for suitable irradiation test capacity and capabilities in the US which cannot be accommodated.

The Spallation Neutron Source (SNS) at Oak Ridge National Laboratory (ORNL) in Tennessee is currently the world's brightest source of pulsed neutrons for scientific research and industrial development and is propelled on its front end by one of the world's highest current proton accelerators [10,11]. This accelerator is further undergoing a planned upgrade which would increase the proton power from 1.4MW to 2.8MW in the next decade [12]. In this article, we show how a facility can be constructed at the SNS which will use less than 1% of this power to address both of the national shortcomings mentioned above with a negligible impact on the existing or future neutron scattering facilities at ORNL. We begin by providing a brief introduction to the technique of µSR and to SEE and recent historical developments in the United States that led to the current design concept. We will follow with design specifics, expected beam characteristics from numerical simulations, and suggestions for an initial facility layout that can be constructed in the next decade.

[*] williamstj@ornl.gov
[+] gmacdoug@illinois.edu

*Muon Spin Rotation/relaxation/resonance*

The technique of µSR involves the implantation of spin-polarized muons into materials of interest and monitoring products of the muon decay to provide extremely sensitive measurements of local magnetic field distributions and fluctuations [13].

To create muons, a proton beam with an energy of at least 330 MeV is directed into a target of a low-Z material (typically carbon or beryllium) to produce unstable pions. The pions decay with a mean lifetime of 26 ns via the weak interaction into a muon (or anti-muon) and an anti-muon neutrino (or muon neutrino). The beam characteristics of the resulting muon depends on the momentum of the pion from which it decays. Positively-charged pions at rest diffuse to the surface of the target and produce so-called "surface muons", which are the most commonly used for µSR experiments. The surface muon is produced with a well-defined kinetic energy of 4.11 MeV and will stop in the bulk of fairly thin samples; with a stopping range of ~150 mg/cm$^2$, a typical stopping depth will lie in the range 0.1-0.3 mm in transition metals. Conversely, muons produced from pions that have been ejected from the target are known as "decay muons". Decay muons have kinetic energies as high as hundreds of MeV and can penetrate more deeply into samples or facilitate the measurement of materials inside high pressure environments or other enclosures. Finally, beams of "low energy muons" can be produced by reducing the momentum of surface muons to the range of 0.5 to 30 keV, currently done successfully at one facility by stopping them in thin films of adsorbed noble gases and reaccelerating them with electric fields [14]. Beams of low-energy muons can be used to measure the properties of thin film samples, nanostructures, surfaces and interfaces, but require a beam with a high initial flux due to the sharp (4-5 orders of magnitude) drop in the muon flux during the slowing process [14].

Because the pion decay is governed by the parity-violating weak interaction, both the neutrinos and anti-muons produced from positive pions are exclusively left-handed; that is, their spin is antiparallel to their linear momentum. Thus, surface muons collected ~90° from the initial proton direction will be 100% spin polarized and easily separated from other particles in the beam. Decay muons have marginally lower (>80%) spin polarization since they are produced and have to be separated in-flight from pions and positron/electrons. Either positive or negative muons can be produced, but since negative muons have more complex interactions with materials, µSR experiments in current facilities heavily favor use of positively charged muons. Though there has been a recent increase in techniques that take advantage of negatively-charged muons, we will exclusively be referring to the positively charged particles when discussing muons in this manuscript. After being implanted into samples, the muon spin direction evolves in a way that reflects material properties, and the particle itself undergoes a weak decay into a positron and two neutrinos. Again, due to parity violation, the positron is emitted preferentially along the final muon spin direction. A µSR experimenter tracks the directional asymmetry of the positron momenta as a function of time to infer information about the ensemble of muon spins, similar to the technique of NMR without the need of large magnetic fields.

µSR has led to important results over the past three decades in the study of magnetism, superconductivity, quantum diffusion, chemistry, semiconductor physics, battery science, and spintronics, among other fields [7]. In the study of magnetism especially, this technique has been repeatedly shown to be highly complementary to neutron scattering: where neutrons provide bulk measurements of correlations in reciprocal space and fluctuations on the timescale of picoseconds (THz), µSR is a local, real-space probe, sensitive to fluctuations on the timescale of microseconds (MHz). There is a high degree of overlap between the user base of the two techniques, and for this reason 3 of the 4 existing large-scale µSR user facilities in the world are co-located with neutron sources (PSI in Switzerland [15], ISIS in the United Kingdom [16,17], and J-PARC in Japan [18]). Additionally, there is a small user program at the MuSIC source in Osaka, Japan [19] and the Chinese Spallation Source is currently planning to build a high-intensity muon beam [20]. The glaring exception is in North America, where the sole muon source is located at the meson accelerator laboratory TRIUMF in Vancouver, Canada [21]. Since the closure of LAMPF in 1995, the United States has not developed a µSR facility.

To address this national shortcoming, there have been several efforts in recent years to assess the feasibility of building a US facility for performing µSR, including a conversation during the construction of the first target station (FTS) of the Spallation Neutron Source (SNS) (2000), during planning of ProjectX at Fermi National Accelerator Laboratory (2013) [22] and the Transformative Hadron Beamlines initiative at Brookhaven National Laboratory (2014) [23]. In each case, these proposals faced technical design challenges that have prevented further progress. In 2016, a workshop was held entitled "Future Muon Source Possibilities at the SNS" [24], at which the ORNL-developed technology at the heart of the current proposal was first discussed in the context of µSR. As presented below, this technology is well-suited to not-only overcome previous limitations, but paves the way for the construction of a pulsed µSR facility in the US with timing resolution approaching the theoretical minimum and beam intensity an order of magnitude higher than current sources. This would truly be a next-generation muon facility and would open up the technique to capabilities never before considered.

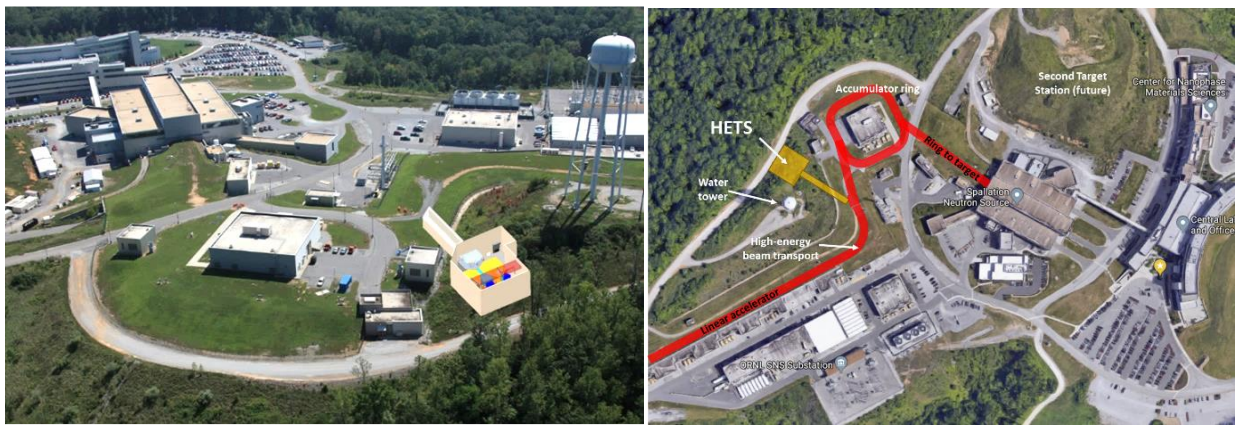

*Figure 1. The High Energy Neutron Test Station (HETS) placed between the SNS site water tower and the accumulator ring.*

*Single Event Effects testing*

Electrical faults known as Single Event Effects (SEE) occur when integrated circuits are exposed to ionizing radiation [25]. Relatively low levels of radiation encountered daily are sufficient to create such events, and they must be taken into account during the design step of modern instrumentation. This is particularly relevant to the aviation industry. Radiation in the Earth's atmosphere originates from cosmic and solar rays and exists naturally from sea level to aircraft altitudes to outer space in varying intensity and type. It is known that the primary radiation type driving SEE in the atmosphere is high-energy neutrons [25, pp. 55]. Thermal neutrons, muons, pions and protons are also of concern to SEE investigators [25, pp. 53].

Modern trends towards automation and increasing complexity in the transportation sector has increased the likelihood that a SEE event will create a failure in a critical system. The increasing number of gate features on integrated circuits, decreasing size, vastly increased memory bits in designs, lower gate voltages, and the use of thermal neutron reactive materials such as boron intimately in circuit fabrication each raise the probability of unplanned events. This is fueling a rise in concern in the avionics and aviation industries.

To deal with this problem, associated standards boards and regulatory agencies are beginning to require SEE assessments for new craft that cite the results of irradiation test data. Changes currently being discussed will result in an increase in demand for irradiation testing in the US which exceed the capacity of existing facilities. The proposed irradiation test facility described below will increase the SEE test capacity by four times, add capability for both device and complete system testing, and allow industrial users to apply high-energy neutrons, thermal neutrons, muons and protons as requested for their certification needs. Fast access for industry users will be a cornerstone feature of the facility's operation, something viewed as essential for rapidly changing commercial businesses.

Ground-based electronics are also vulnerable to SEE, even though the neutron flux reaching the earth's surface is a few hundred times less than for aircraft flying at 40,000 feet [26]. Susceptibility trends for micro-electronics in ground-based systems move even faster in terms of sophistication and growing applications on critical / safety related systems, e.g., the emerging use of autonomous vehicles. The proposed SEEMS facility will satisfy industry needs to certify critical electronic equipment when appropriate component cross-section data is not available. In addition to small and intense neutron beams for component testing, SEEMS will support testing of entire electronic systems with large beams up to $1\times2$ m$^2$ – a capability not currently available in the US.

**2. Facility Concept Overview**

Both the next-generation µSR and high throughput SEE testing facility referred to above depend on the redirection of a small percentage of the 1.3 GeV protons from the linear accelerator arm of the SNS towards a suitable target. Key to our design is pioneering work performed at ORNL in developing laser stripping technology [27] that makes possible the removal of short, intense proton pulses from the primary beam, while leaving the majority of the beam intact for subsequent production of neutrons. In contrast to the parasitic target strategy pursued by other muon sources, our concept separates the muon and irradiation facilities from the neutron production both geographically and in the timing of the sources, resulting in a negligible impact on the primary neutron scattering mission of the SNS.

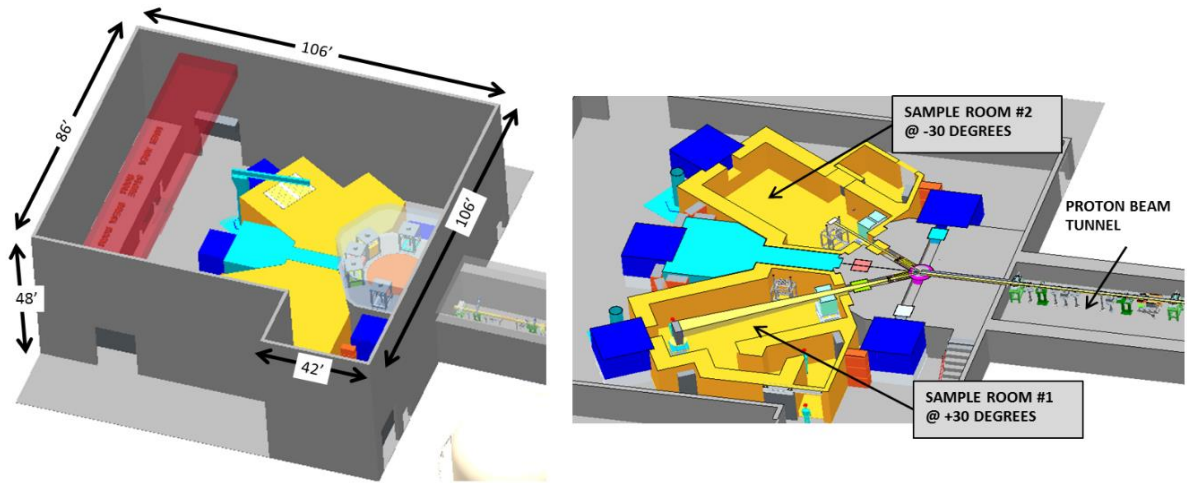

*Figure 2. The High Energy Neutron Test Station layout shows two neutron test areas (sample rooms) at +/- 30° off the direction of the incident proton beam. Figured adapted from [28].*

Moreover, the existing resources at the SNS and ORNL provide the means to construct this facility for a fraction of the cost of a new dedicated accelerator facility, on a much shorter timescale, and with a lower ultimate operational cost by taking advantage of the existing user facilities.

The SNS operates as a US Department of Energy, Basic Energy Sciences User Facility for neutron scattering science; it has staff with experience accustomed to visiting researchers bringing samples for testing in neutron beams. Advantages of co-locating this facility at the SNS include:

- The SNS is a relatively new accelerator facility with at least another 50 years of future life;
- Extensive beam operation / production: about 5000 hours (200 days) of production is scheduled each year;
- SNS has high reliability and stability, with operational availability > 90%;
- Staff have expertise in radiation detection and safety;
- An ongoing upgrade project to the SNS facility will increase proton energy to 1.3 GeV and double the accelerator's power to 2.8 MW [12].

The discussions that led to the conceptual facility discussed here began with a 2014 study funded by the Federal Aviation Administration (FAA) for an SNS-based SEE test facility at the ORNL to provide neutron irradiation capabilities for avionics (or ground based) devices and systems [28]. This study considered placing a SEE testing beamline at the existing SNS neutron target or at the future Second Target Station. These were less desirable since they viewed a neutron moderator and had space limitations. The most capable option that emerged from that study used laser stripping, described below, to divert a small portion of the primary beam from the SNS accelerator to a new spallation target station with two SEE test areas where generated neutron beams would be delivered. Called the High Energy Neutron Test Station (HETS) [28], the design could provide neutron beams with an energy spectrum matching that in the atmosphere, but at higher intensities needed to accelerate the determination of event rates in practical timeframes. Intensities and neutron beam spot size could be varied for either small devices or large systems. Thermal neutrons could be added as needed by movable moderators. Protons scattered off the target would normally be deflected from the test area by magnets but could be disabled when desired to mix with the neutron beam. HETS was envisioned to be placed between the existing water tower and the SNS accumulator ring as shown in Figure 1. It would divert a small amount of the proton beam extracted from the end of the high-energy beam transport section of the accelerator, and before the junction with the accumulator ring. The HETS target station concept (Figure 1) had space and functional capacity to support other missions that were not covered in the FAA study. For example, space-based electronics experience SEE from radiation consisting of primarily (~90%) of high-energy protons. Protons alone could be sent to a third test area located between the neutron test areas (i.e., into the light blue region at 0° in Figure 2), with proton energies ranging up to 1.3 GeV. Two additional beam ports were also envisioned to be placed at ±90º to the incident proton beam for unspecified use in HETS.

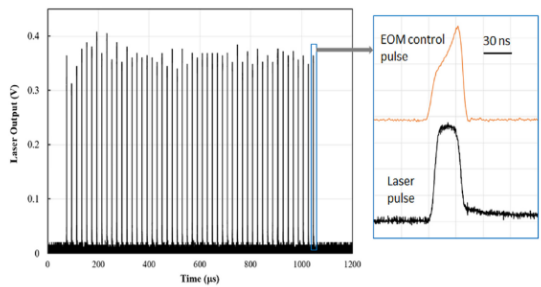

*Figure 4. The measured proton pulses generated by the 30ns/30kHz laser pulse. The resulting proton pulse retained the pulse width and frequency and allowed calculation of the stripping efficiency. Reproduced with permission from Nucl. Inst. Meth. A. 962, 163706 (2020). Copyright 2020 Elsevier.*

In 2016, a workshop was convened to discuss the possibility of co-locating a muon source at the SNS for the purposes of performing μSR experiments – a technique strongly complimentary to the existing neutron scattering program [24]. First considered was the placement of a parasitic muon target upstream from the existing neutron spallation target – the primary strategy employed at other neutron facilities with co-located muon sources [15,16,18,19]. Though this mode of production in a future muon facility would benefit from an established design approach with known source physics parameters, it was noted that it comes with significant complications for the concurrent neutron scattering program. At PSI, for example, the entire proton beam (ca. 1 MW) passes through the muon target on its way toward the neutron target. About 2% of the proton energy is deposited in the muon target. However, due to proton beam scattering from the upstream muon target, the neutron intensity out of the primary target is reduced by about 30% [29], though this varies significantly for other co-located facilities.

In this context, μSR experts at the 2016 workshop quickly realized the benefits of the HETS proposal outlined above and developed the groundwork for the current source design. Utilizing laser stripping, an ideal pulsed proton beam could be selected from the linear accelerator and diverted to a separate muon target, resulting in what would easily be the highest flux and highest resolution pulsed μSR source in the world. As this source option makes use of protons prior to the SNS accumulator ring, there would be no impact on the primary neutron target. The SEE / muon facility would be physically well-separated from the neutron scattering science building, thus avoiding any background issues generated from the muon target affecting sensitive neutron instruments.

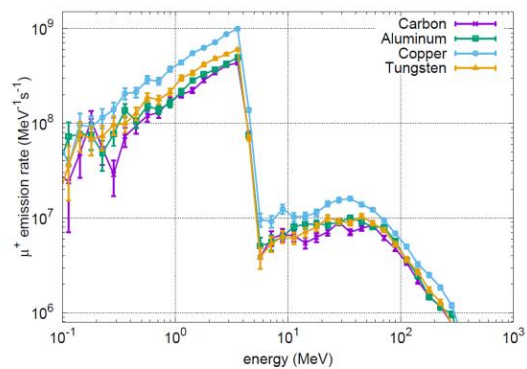

*Figure 3. Results of the Monte Carlo simulations for 5 mm wide copper and tungsten muon production targets. This shows the number of muons of a given energy that result only from pions decaying at rest in the target.*

Fortuitously, surface muons are extracted most efficiently at ±90º to the incident proton beam at the muon target, meaning that the μSR beamlines could be incorporated into the existing HETS facility design using the available perpendicular beam ports. This realization yielded the central elements of the SEEMS design concept described in this report, allowing for the construction of a single facility with strong scientific and industrial use missions, again leveraging the construction and operational costs to serve both communities. The SNS accelerator total power can easily be increased to entirely compensate for the small (~0.2%) proton current directed towards SEEMS. Below we outline the key components and technologies of the SEEMS facility design.

## 3. Laser Stripping

The core technological development for this proposed facility is the use of a laser to strip electrons from the H$^-$ beam of the SNS linear accelerator (linac) to create an optimized proton pulse which could be directed towards a μSR/SEE facility. The primary H$^-$ beam is characterized by a complicated timing structure, wherein a series of micro-pulses of 140 ps repeat at 402.5 MHz and are bunched into mini-pulses of 645 ns repeating at 1.06 MHz, which are then in turn bunched into macro-pulses of 1.0 ms at 60 Hz [30]. The ability to use a laser pulse to strip electrons from this beam had been successfully demonstrated prior to this facility concept, without any modification to the intrinsic time structure [30]. These previous measurements were carried out at the SNS using a laser transport line that terminated in the High-Energy Beam Transport area of the SNS linac where it interacted with the H$^-$ beam. The neutral H atoms were unaffected by electrostatic fields bending the primary H$^-$ beam into the accumulator ring and continued straight into the linac dump. At that point, a diagnostics test stand measured the dimensions and current (and hence the laser stripping efficiency) of the resulting beam. In the planned SEEMS facility, the neutral H atoms will be stripped of remaining electrons to create the necessary proton beam to its dedicated target.

To extend this work and demonstrate that laser stripping is suitable for the proposed SEEMS facility, further proof-of-principle experiments were performed in 2019 with the laser system modified to operate with a modified time structure idealized for muon production. A 30 ns wide pulse was chosen to be compatible with the pion lifetime, which presents a theoretical minimum for this parameter. A repetition frequency of 50 kHz was also chosen to allow 20 μs between pulses, which is similar to the parameters at other pulsed muon facilities. By measuring the stripping efficiency of the existing laser system with the modified time structure, it was possible to extrapolate the required laser power to achieve a high stripping efficiency on the entire H$^-$ beam in a full facility concept. The resultant data demonstrate that the laser stripping system at the SNS can reliably operate continuously with the pulse width and frequency required for the ultimate SEEMS facility (see Figure 3) [31].

These were low-power experiments, but a high laser power is required to achieve high stripping efficiencies. A final design will likely incorporate an optical cavity, which may slightly increase both the pulse width and the effective laser duty cycle. Preliminary calculations suggested that a feasible design would create ~70% laser efficiency with a laser duty cycle of 0.225% (45 ns pulse at 50 kHz). With the SNS accelerator power of 2.8 MW, this would give the beam power at the muon target of (2.8 MW) x (0.225%) x (70%) = 4.41 kW. This would produce a muon pulse with a FWHM of ~50 ns [24]. The muon pulse width could be varied by chopping the beam or by modifying the length of the laser pulse through conventional means. This is a major advantage of the current technique, as the flux and pulse width can be varied facility-wide at any time by changing the laser pulse structure. The above-mentioned efficiency measurements refined the preliminary estimates and show that the use of existing high-power lasers and optical cavities could achieve 90% stripping efficiency with a muon pulse width of 49.7ns [31]. The success of this project addresses one of the major sources of risk with the design of the SEEMS facility.

| Target Material | Width (mm) | Surface Muon Emission* ($\mu^+$ / sec) |
|---|---|---|
| Carbon | 2.0 | 1.37 x 10$^9$ |
|  | 3.0 | 1.29 x 10$^9$ |
|  | 4.0 | 1.19 x 10$^9$ |
|  | 5.0 | 1.14 x 10$^9$ |
|  | 6.0 | 1.08 x 10$^9$ |
|  | 7.0 | 1.04 x 10$^9$ |
| Aluminum | 2.0 | 1.31 x 10$^9$ |
|  | 3.0 | 1.33 x 10$^9$ |
|  | 4.0 | 1.33 x 10$^9$ |
|  | 5.0 | 1.27 x 10$^9$ |
|  | 6.0 | 1.24 x 10$^9$ |
|  | 7.0 | 1.22 x 10$^9$ |
| Copper | 2.0 | 1.86 x 10$^9$ |
|  | 3.0 | 2.21 x 10$^9$ |
|  | 4.0 | 2.37 x 10$^9$ |
|  | 5.0 | 2.52 x 10$^9$ |
|  | 6.0 | 2.58 x 10$^9$ |
|  | 7.0 | 2.61 x 10$^9$ |
| Tungsten | 2.0 | 0.903 x 10$^9$ |
|  | 3.0 | 1.13 x 10$^9$ |
|  | 4.0 | 1.37 x 10$^9$ |
|  | 5.0 | 1.58 x 10$^9$ |
|  | 6.0 | 1.72 x 10$^9$ |
|  | 7.0 | 1.82 x 10$^9$ |

*Surface muon emission in the simulations corresponds to muons created from pions that decay at rest in the target and where those muons are ejected from the target with a kinetic energy less than 4.11 MeV. The simulations do not include pion diffusion within the target.*

*Table 1. Results of the target simulations for various target materials and widths. This shows that the target material and geometry can be optimized to produce a facility total muon flux of at least 10$^9$ surface muons per second.*

## 4. Target

Of the four largest μSR sources currently in operation, three are co-located with a source for neutron scattering science. In all cases, the muon target is upstream from the neutron target with the majority of protons passing through and only a small percentage (<5%) of the primary beam contributing to muon production. An additional advantage of the proposed SEEMS facility is that the target design can be optimized for muon and neutron production in several closely spaced configurations and materials without need to preserve beam current for subsequent production activities (e.g. a neutron scattering source) [32]. To facilitate the two central missions of a SEEMS facility, the target could be a hybrid of two material types arranged in series or the muon target might have a central passage allowing full energy to a downstream neutron target, presumed to be tungsten. Power deposited into the target necessitates an efficient design for removing heat and prolonging the lifetime of the targets. Coupled with the target requirement for efficient muon production, there is a wide range of target geometries and compositions that are currently being explored. Required power and target design will dictate shielding requirements around the target monolith.

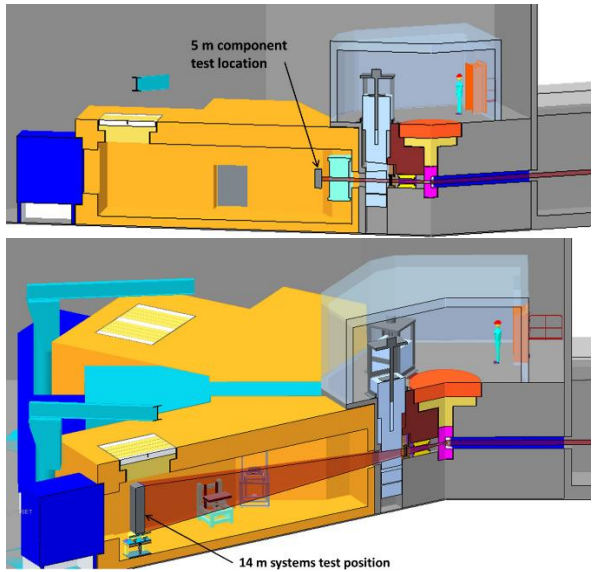

*Figure 5. Cross-sectional views of the HETS target monolith and test enclosures.*

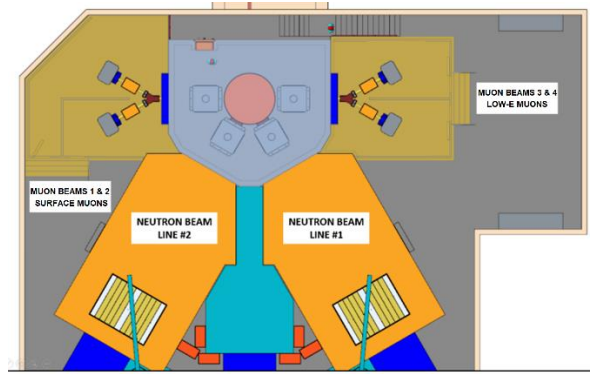

*Figure 6. Overhead schematic showing the potential layout of the combined µSR/SEE facility. Components are for illustration purposes only, and not to scale.*

Monte Carlo physics simulations have been performed with the MCNP6.2 code [33] to explore the size, geometry, and material properties of an efficient muon target design. Conventional design choices such as flat plates made of graphite or beryllium were considered, as well as more innovative designs. These include targets made of diamond, or alternate geometries such as a series of stacked foils or longer, thicker targets- both aimed at maximizing surface area for surface pion creation [34]. Preliminary simulations focused on flat targets of length 100 mm and using a proton beam with zero width (pencil beam). Simulations used varying widths of the plate in the direction perpendicular to the incoming proton beam. These simulations also looked at various target materials, namely carbon, aluminum, copper, and tungsten. A sample of these results is shown in Figure 4, and the resulting muon flux for the various target materials and geometries is shown in Table 1. These simulations calculate the muons created in the target from pion decay that are emitted from the target (the muons do not stop in the target). This simulation does not include pion diffusion, so it will undercount the number of surface muons. However, it is a reasonable approximation of the expected muon current emitted from the target. This data shows a sharp drop at 4.11 MeV, above which there are no surface muons, but shows the large number of decay muons produced due to the high incident proton energy. The highest muon flux was obtained from the copper target, followed closely by tungsten. This is not a usual muon target material, due to the increased production of neutrons, which would be an additional background for neutron instruments in facilities where muon targets are co-located with neutron scattering science instruments [32]. However, the SEEMS facility would be well-separated from the neutron scattering science experiment hall, making this consideration less relevant. The higher neutron production would also lead to higher radiological conditions around the target; however, the lower beam power would reduce the required shielding and it is envisioned that the target can be serviced via overhead crane access. Monte Carlo simulations assuming a tungsten target indicated that the required shielding would be less than used elsewhere at the SNS. Tungsten ended up being a nearly optimal choice of target material, and the calculated generation of surface muons for a 5 mm wide copper target was approximately $10^9$ µ$^+$/sec (shown in Figure 4), which is an order of magnitude higher than any currently existing pulsed µSR facility. The inferred ratio of approximately $6\times10^{-5}$ µ$^+$/p$^+$ is consistent with proton-pion-muon conversion calculations for other facilities [32, 35], where the generated flux of surface muons is also on the order of $10^{-5}$ µ$^+$/p$^+$. Proton power to the SEEMS target would be less than 5 kW, an important threshold for radiological hazards, and requires less shielding than for other target systems at the SNS.

The simulations also highlight that approximately 10 times more decay muons are produced than surface muons due to the high proton energy from the SNS linac

(1.3 GeV after the currently ongoing accelerator upgrade project). In light conventional muon spectrometer designs and the overwhelming desire for low-energy muon production, future design activities will focus on maximizing the production of surface muons. Future studies may also consider less conventional target geometries, as well as other target materials, such as nickel. Tungsten is the preferred choice for the SEE neutron target however, and simplifying the target design by using only tungsten for both muon and neutron production may be preferred.

Whereas the original beam power for the HETS facility was envisioned as less than 3 kW, the increased power for the planned muon production is still under 5 kW. This change is small enough that target cooling foreseen in HETS with helium gas can likely be maintained. The small power increase avoids additional target monolith shielding as well, and avoids the added cost associated with more shielding or changing to water cooling (see Figure 5).

## 5. Beamlines

Using the existing building model from the HETS study, a concept for muon experimental areas was developed by adding a primitive layout for 4 µSR beamlines, two each on the ±90º beam ports of the HETS concept. The layout is shown in Figure 6.

The previous proposal for the HETS included two identical testing stations receiving high-energy neutron beams at the ±30° beamline positions. These stations are unchanged in the SEEMS concept, with each 9×3 m (interior area) testing stations offering two irradiation positions, one for device testing and one for system testing. The position for component irradiation will be located at about 5 m from the target, and the beam line will provide fluxes of 10 MeV or greater neutrons of up to $10^7$ n/cm$^2$/s onto areas as large as $20 \times 20$ cm$^2$. The position for system irradiation will be located at the far-target position (14 m from the target). At the system position, it will be possible to deliver peak above-10 MeV fluxes up to $1.3\times10^6$ n/cm$^2$/s over an area of 0.56 $\times$ 0.56 m$^2$ (propagation of component testing beam to the back of enclosure). Alternately, it will be possible to deliver a beam over an area of $1\times2$ m$^2$ with a peak above-10-MeV fluxes up to $2\times10^5$ n/cm$^2$/s. At these flux levels and beam dimensions, the peak integral in-beam neutron currents are equivalent for all irradiation conditions. The neutron pulse timing is not a factor in these measurements, as they are concerned only with total dose accumulation and the energy profile. Figure 5 shows cross sectional views of the test enclosures.

As a neutron beam enters a test area, real-time diagnostic equipment will determine its flux intensity to provide a means to tailor the beam to user requirements and to quantify the delivered fluence. The neutron beam monitoring and diagnostics are envisioned using a three-part approach. Neutron energy spectra and absolute intensity will be measured by activation foils and proton recoil telescopes during dedicated calibration periods, during which real-time detectors will be calibrated to provide scaling estimates for the established spectral distributions. Active calibration will be provided by a proton-recoil telescope of 1 cm$^2$ rastered across the active area. For passive calibration and monitoring, activation foil packets will be employed that will be assayed following irradiation. For real-time monitoring, transistor arrays and fission chambers will be deployed.

The ±90º beam ports will extract high numbers of surface muons, which will comprise the core instruments of the initial µSR facility. Surface muon beams are the most common in current µSR, due to the ease of extraction and the high level of spin polarization. Surface muons originate from pions decaying at rest on the surface of the target, implying a long, thin target can be used and the muons will be emitted in all directions. The choice to extract beam from the ±90° positions minimizes contamination from other particles in the primary beam. Beamlines using surface muons are typically built to be short to minimize spatial broadening of the muon pulse and to minimize decay of the muons in flight. As charged particles, the muons are easily steered and filtered in the beamline using multipole magnets, and readily stopped in samples of interest. Typical stopping distances are 0.1-0.3 mm for surface muon beams in metals, allowing small samples to be easily measured in a typical µSR experiment. As it is a dedicated muon production target, it is feasible that 25% of the surface muons could be directed into the muon beamlines. Our initial design uses half of these surface muons to maintain two workhouse µSR instruments for the characterization of bulk material properties. This is akin to instruments in other facilities, though the high muon flux offered by the favored source design will increase throughput and allow measurement of smaller samples and more complex measurement apparatuses. Alternatively, the laser stripping technology would allow the one to vary the pulse structure, increase the time between muon pulses, and allow the measurement of smaller magnetic moments and long timescale phenomena.

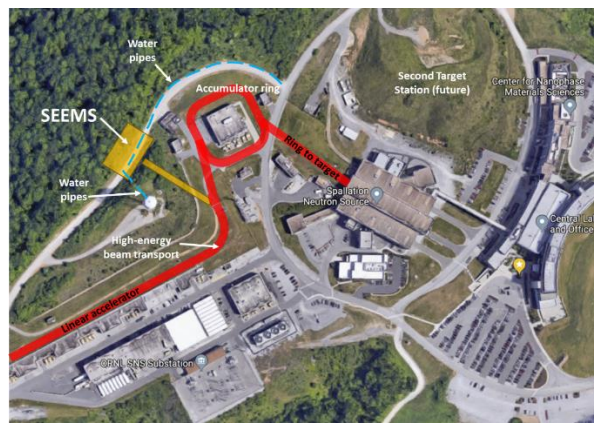

*Figure 7. The proposed location of the SEEMS facility (orange) relative to the accelerator beam (red) and other existing and planned infrastructure. An enlarged SEEMS building would be moved further away from the accelerator compared to the HETS concept.*

Significantly however, when operating in high a flux configuration, the facility will also be capable of producing large numbers of low energy muons (LEM), with kinetic energies of 1-30 keV. Muons with smaller momenta will have shorter penetration depths into a sample than typical for surface muons. This is advantageous for studying thin film samples or multi-layer devices. Kinetic energies at this level can be tuned with applied electric fields, further allowing for depth resolved studies involving surface effects or interfaces. With enough beam current, this depth resolution can be combined with muon focussing and beam steering to create a novel muon microscope [36], capable of mapping out local magnetic fields inside samples of interest in three dimensions.

LEM beams can be created from surface muons, which are then slowed to the required energy [37,38]. The high flux of the current source design more than makes up for the large beam current losses associated with the slowing processes in the creation of LEM beams. Our initial facility design dedicates half of the surface muon production to the creation of LEM capabilities. Future design work will be dedicated to advanced beamline planning that would ultimately determine flux and beam characteristics at the position of the planned instruments, as well as decisions regarding the most efficient means of extracting the muon beam from the target (e.g. through the use of dipole magnets around/near the target).

Though not in the layout presented in Figure 6, it would also be a simple matter to include a decay muon channel downstream from the initial target. Decay muons are created from pions that are ejected from the target and consequently have greater momenta. They penetrate farther into samples, and thus can be used to measure materials that are shielded behind capping layers, liquids inside containers, or crystals in pressure cells or other unique environments.

The diverse scientific profile offered by these three types of µSR beamlines encourages us to include all three types in the proposed facility. Furthermore, having flexibility of the beamlines to perform multiple types of experiments with a wide range of muon momenta will provide the greatest flexibility to address current and future scientific problems.

## 6. SEEMS Location

The baseline location for the SEEMS facility and its building footprint are the same as for the HETS concept. This location extracted the proton beam just before the accumulator ring, placing the target building north of the linac. This offers a convenient location to electrostatically extract the portion of the beam ionized by the laser pulse.

The required space for muon beamlines is less well defined than for the SEE test areas, and it is recognized that greater space might be needed than what is available within the baseline concept's building footprint. Locating a facility on the SNS site should consider a suitable connection to the accelerator for laser stripping and proton beam extraction, minimizing rework of existing underground infrastructure, and the local terrain. The baseline location and facility size satisfied these considerations for the original HETS mission, but it is bounded by the site water tower and accumulator ring support buildings.

If more space is needed for muon beamlines, the HETS building size and position could be modified for an enlarged SEEMS facility. Widening the building would require moving it further north and across the access road as illustrated in Figure 7. This would cross existing underground pipes to the water tower that follow along and cross the road. The terrain north of road also begins to slope downward more steeply. In addition to the expense of a larger building, the cost for reworking the water pipes and additional terrain modification would have to be considered.

## 7. Conclusion

This manuscript has outlined the preliminary work done to realize a Single-Event Effects and Muon Spectroscopy (SEEMS) facility at the SNS. While only at the initial stages of design, it highlights the scientific opportunities presented by such a facility. Not only does it fill a crucial gap in US SEE testing needs, but it provides US researchers with a domestic source for µSR measurements. In both respects, the SEEMS facility can offer capabilities not present anywhere else in the world. It also expands the µSR capabilities in North America, being complimentary to the continuous muon source at TRIUMF. By leveraging the existing accelerator and user facility infrastructure at the SNS and ORNL, this can be done for a fraction of the cost of a green field project.

The user communities have continued to be engaged in this process and will be crucial to the development of the concept moving forward. The next steps will involve further development of the concept and refining the facility parameters. This will allow for better bounds on the timeline and cost for the facility.

**Acknowledgements**

A portion of this work used resources at the Spallation Neutron Source, a DOE Office of Science User Facility operated by the Oak Ridge National Laboratory. This manuscript has been authored by UT-Battelle, LLC under Contract No. DE-AC05-00OR22725 with the U.S. Department of Energy. The United States Government retains and the publisher, by accepting the article for publication, acknowledges that the United States Government retains a non-exclusive, paid-up, irrevocable, world-wide license to publish or reproduce the published form of this manuscript, or allow others to do so, for United States Government purposes. The Department of Energy will provide public access to these results of federally sponsored research in accordance with the DOE Public Access Plan (http://energy.gov/downloads/doe-public-access-plan).